\documentclass[11pt,psfig]{article}

\usepackage{amsfonts,amssymb,amsmath}
\usepackage{natbib}
\usepackage{epsfig}
\usepackage{bm}
\usepackage{multirow}
\usepackage{ulem}   % Added by Bryan

\newcommand{\bl}{\begin{flushleft}}
\newcommand{\el}{\end{flushleft}}

\newcommand{\bc}{\begin{center}}
\newcommand{\ec}{\end{center}}

\begin{document}

\baselineskip=24pt

\title{Fitting Semiparametric Cumulative Probability Models for Big Data}
\author{Chun Li, Guo Chen, Bryan E. Shepherd}
\maketitle

\section*{Abstract}

Cumulative probability models (CPMs) are a robust alternative to linear
models for continuous outcomes.  However, they are not feasible for very
large datasets due to elevated running time and memory usage, which depend on
the sample size, the number of predictors, and the number of distinct
outcomes.  We describe three approaches to address this problem.  In the
divide-and-combine approach, we divide the data into subsets, fit a CPM to
each subset, and then aggregate the information.  In the binning and rounding
approaches, the outcome variable is redefined to have a greatly reduced
number of distinct values.  We consider rounding to a decimal place and
rounding to significant digits, both with a refinement step to help achieve
the desired number of distinct outcomes.  We show with simulations that these
approaches perform well and their parameter estimates are consistent.  We
investigate how running time and peak memory usage are influenced by the
sample size, the number of distinct outcomes, and the number of predictors.
As an illustration, we apply the approaches to a large publicly available
dataset investigating matrix multiplication runtime with nearly one million
observations.

Keywords: Cumulative probability models; semiparametric; big data

\newpage
\section{Introduction}

Continuous outcomes often require a transformation prior to fitting linear
models.  The choice of transformation is not always clear, and different
transformations may result in different conclusions.  Semiparametric linear
transformation models (Zeng and Lin, 2007) assume a linear model after a
monotonic transformation which is nonparametrically estimated.  The outcome
and predictor variables are linked through an unobserved latent variable,
where the latent variable is connected to the predictors as in a traditional
linear model and to the outcome through an unknown monotonic transformation.
These models are desirable because they require neither an explicit
transformation of the outcome nor a model for the conditional expectation.
Instead, the conditional distribution given covariate values is modeled, from
which conditional expectations, quantiles, and other quantities can be
derived.

Recently, Liu et al. (2017) showed that semiparametric linear transformation
models can be fit using cumulative probability models (CPMs).  CPMs have
typically been reserved for the analysis of ordered categorical response
variables; e.g., proportional odds and other ``cumulative link models''
(McCullagh 1980; Agresti 2010).  With CPMs, the continuous outcome variable
is effectively treated as if it were ordinal because what matters is the
order of the outcome values, not the values themselves.  Liu et al. (2017)
showed that the multinomial likelihood used for CPMs is 
the likelihood of a semiparametric linear transformation model.  Using
computer simulations, they showed that CPMs perform well in many scenarios.
Under mild conditions, CPMs yield estimates that are consistent and
asymptotically normal (Li et al. 2021).

Because the transformation between the outcome and the latent variable is
modeled nonparametrically, CPMs can be slow to fit with large samples,
especially when there are many unique outcome values (Liu et al., 2017).  The
sparse nature of the score and Hessian matrices of CPMs can be exploited to
make computation feasible for sample sizes in the thousands (Harrell, 2020).
However, even when employing computationally efficient algorithms, CPMs are
not able to handle larger sample sizes.  For example, we cannot fit a CPM to
a simulated dataset with 50,000 distinct outcomes on a server with 48 cores
and 192 gigabytes of memory.  The robustness and flexibility of CPMs make
them desirable for analyses of large datasets; however, current big data
implementations are not feasible.

The purpose of this paper is to describe and evaluate methods for fitting
CPMs for big data, either by reducing the sample size through dividing the
data into subsets or by reducing the number of distinct outcomes via binning
or rounding.  In Section 2, we introduce three approaches to fitting a CPM
for a large dataset.  In the divide-and-combine approach, the sample size of
each individual CPM is greatly reduced.  In the binning and rounding
approaches, the outcome variable is redefined to have a greatly reduced
number of distinct values.  We carry out computer simulations to evaluate and
compare these approaches in Section 3, and apply them to a real data example
in Section 4.  Section 5 contains a discussion.

\section{Methods}
\subsection{Cumulative probability models}

In this subsection we briefly describe cumulative probability models (CPMs)
and relevant notation; details can be found in Liu et al. (2017).  One way to
motivate CPMs for a countinuous outcome $Y$ is through a linear
transformation model,
\begin{equation}
  Y=H(\bm\beta^T X+\epsilon),
\end{equation}
where $X$ is a vector of $p$ predictors, $\epsilon\sim F_\epsilon$ with
$F_\epsilon$ known, and the transformation $H(\cdot)$ is assumed to be
non-decreasing and otherwise unknown.  It is easy to show that for any $y$,
\begin{equation*}
  \Pr(Y\le y\vert X) =\Pr\{\epsilon\le H^{-1}(y)-\bm\beta^T X\}
  =F_\epsilon\{H^{-1}(y)-\bm\beta^T X\}.
\end{equation*}
Let $G=F_\epsilon^{-1}$ and $\alpha=H^{-1}$.  Then model (1) becomes a CPM,
\begin{equation}
  G\{\Pr(Y\le y\vert X)\} =\alpha(y)- \bm\beta^T X, \text{ for any } y,
\end{equation}
where $G(\cdot)$ is a link function.  The ``intercept'' $\alpha(y)$ in the
CPM has a nice interpretation: It is a transformation of the outcome such
that the transformed value is related to the predictors linearly.  This is
because model (1) can be alternatively expressed as
$\alpha(Y)=\bm\beta^T X+\epsilon$.

Now consider a dataset $\{(y_i,x_i): i=1,\ldots,N\}$, where $N$ is the sample
size.  We first consider the scenario where there are no ties in the
outcomes.  The CPM in (2) becomes
\begin{equation}
  G\{\Pr(Y\le y_i\vert X)\} =\alpha_i-\bm\beta^T X, \text{ for every } y_i,
\end{equation}
where $\alpha_i=\alpha(y_i)$.  Without loss of generality, we assume
$y_1< y_2<\cdots<y_N$; then $\alpha_1<\alpha_2<\cdots<\alpha_N$.  Model (3)
is identical to some models for ordered categorical outcomes: for example,
proportional odds model (when $F_\epsilon$ is the logistic distribution) and
probit model (when $F_\epsilon$ is the normal distribution).  It becomes the
proportional hazards model when $F_\epsilon$ is the left-skewed Gumbel
distribution.

As described in Liu et al. (2017),  the nonparametric
likelihood for the model is 
\begin{equation*}
  L^*(\bm\beta, \bm\alpha) =\prod_{i=1}^N \{G^{-1}(\alpha_{i}-\bm\beta^T x_i) - G^{-1}(\alpha_{i-1} -\bm\beta^T x_i)\}.
\end{equation*}
Here $\alpha_0$ is an auxiliary parameter and $\alpha_0<\alpha_1$.  Because
$L^*(\bm\beta, \bm\alpha)$ is maximized when $\hat{\alpha}_0=-\infty$ and
$\hat{\alpha}_N= +\infty$, it can be simplified as
\begin{equation}
  \{G^{-1}(\alpha_1 - \bm\beta^T x_1)\}
  \{G^{-1}(\alpha_2 - \bm\beta^T x_2) - G^{-1}(\alpha_1 - \bm\beta^T x_2)\} \cdots
  \{1-G^{-1} (\alpha_{N-1} - \bm\beta^T x_N )\}.
\end{equation}
This is the same as the likelihood when the outcomes are treated as if they
were ordered categorical.  We can obtain the nonparametric maximum likelihood
estimates (NPMLE) $(\widehat{\bm\alpha}, \widehat{\bm\beta})$ from this
likelihood.  Then the NPMLE of the transformation function $\alpha(y)$ is
$\widehat{\bm{\alpha}} =\{\hat\alpha_1,...,\hat\alpha_{N-1}\}$.  It is an
increasing step function, with a step for each of the $N-1$ intervals between
adjacent outcome values from $y_1$ to $y_N$.

When there are ties in the outcomes, the above derivations still apply with a
slight modification.  Here model (3) is applicable for every distinct outcome
value, and for each such value, there is a corresponding $\alpha$ value.  Let
$M$ be the number of distinct outcome values.  The NPMLE of $\alpha(y)$
becomes $\widehat{\bm{\alpha}} =\{\hat\alpha_1,...,\hat\alpha_{M-1}\}$, an
increasing step function with a step for each of the $M-1$ intervals between
adjacent distinct outcome values.

Using modern empirical process theory and under mild regularity conditions, Li et al. (2021) showed the consistency and asymptotic normality of the NPMLEs, $(\widehat{\bm\alpha}, \widehat{\bm\beta})$. Asymptotic theory relies on compactness of the estimator of $\alpha(\cdot)$, which does not hold if $Y$ is unbounded. Hence, Li et al (2021) specify lower and upper bounds for $Y$, $L$ and $U$ respectively, such that any observation outside the the interval $(L, U)$ is censored. The corresponding censored data likelihood is equivalent to a likelihood that treats these values as belonging to the lowest / highest ordered categories. Under this censoring, $(\widehat{\bm{\alpha}}, \widehat{\bm{\beta}})$ is consistent for $(\alpha(y), \bm{\beta})$ for $y \in [L,U]$. In addition, $\sqrt{N}(\widehat{\bm{\alpha}}-\alpha(y), \widehat{\bm{\beta}}-{\bm{\beta}})$ converges weakly to a tight Gaussian process whose variance can be estimated as the inverse of the Hessian matrix. Furthermore, the asymptotic variance of  $\sqrt{N}(\widehat{\bm{\beta}}-{\bm{\beta}})$ attains the semiparametric efficiency bound. Details are in Li et al. (2021).

The Hessian matrix has dimensions $(M-1+p)\times(M-1+p)$.  Because of the
special structure of the likelihood function (4), the portion of the Hessian
matrix with respect to the alpha parameters is tridiagonal.  This allows
matrix inversion efficiently through Cholesky decomposition (Sall 1991).
Taking advantage of these facts, Harrell implemented a computationally
efficient algorithm in the \texttt{orm()} function in the \texttt{rms} R
package (Harrell 2020) to fit CPMs.  However, as the number of parameters
increases with the number of distinct outcome values, the function eventually
fails for very large datasets due to heavy demand on computation time and
memory usage.  The next three subsections describe approaches to fit CPMs for
very large datasets.

\subsection{Divide and combine}

In the divide-and-combine approach, the data are evenly divided into subsets,
a CPM is fit to each subset, followed by a final step to aggregate all the
information.  The goal is to make the sample sizes of the subsets small
enough to fit CPMs relatively quickly with a reasonable amount of computer
resources.  Let $K$ be a target number of subsets, and $n_k$ be the size of
subset $k$ ($k=1,\cdots,K$).  If $N$ is a multiple of $K$ then
$n_1=\cdots=n_K=N/K$.  If not, our implementation ensures that
$\max|n_{k}-n_{k'}|\le1$ for any two subsets.  Let $m_k$ be the number of
unique outcome values in subset $k$.  Let
$c^{(k)} =(\widehat{\bm\alpha}^{(k)}, \widehat{\bm\beta}^{(k)})$ be the
vector of parameter estimates from subset $k$, where
$\widehat{\bm\alpha}^{(k)}$ has length $m_k-1$ and $\widehat{\bm\beta}^{(k)}$
has length $p$.  Let $c_j^{(k)}$ be the $j$-th element of $c^{(k)}$.

As $\hat\alpha(y)$ is estimated nonparametrically, it is defined only over
the range of outcomes in the data used to fit a CPM.  To ensure that
$\hat\alpha(y)$ is available over the widest range of outcomes from all the
subsets, in the partition step, we randomly assign the $K$ observations that
have the smallest outcome values one to each subset, and similarly, randomly
assign the $K$ observations that have the largest outcome values one to each
subset.  The rest are allocated to the subsets randomly.  This way,
$\hat\alpha(y)$ will be available from all the subsets over the interval
$[y_{(K)}, y_{(N-K)}]$, spanning from the $K$-th smallest to the $K$-th
largest outcome values.

\begin{figure}[!tb]
  \begin{center}
    \includegraphics[width=6in]{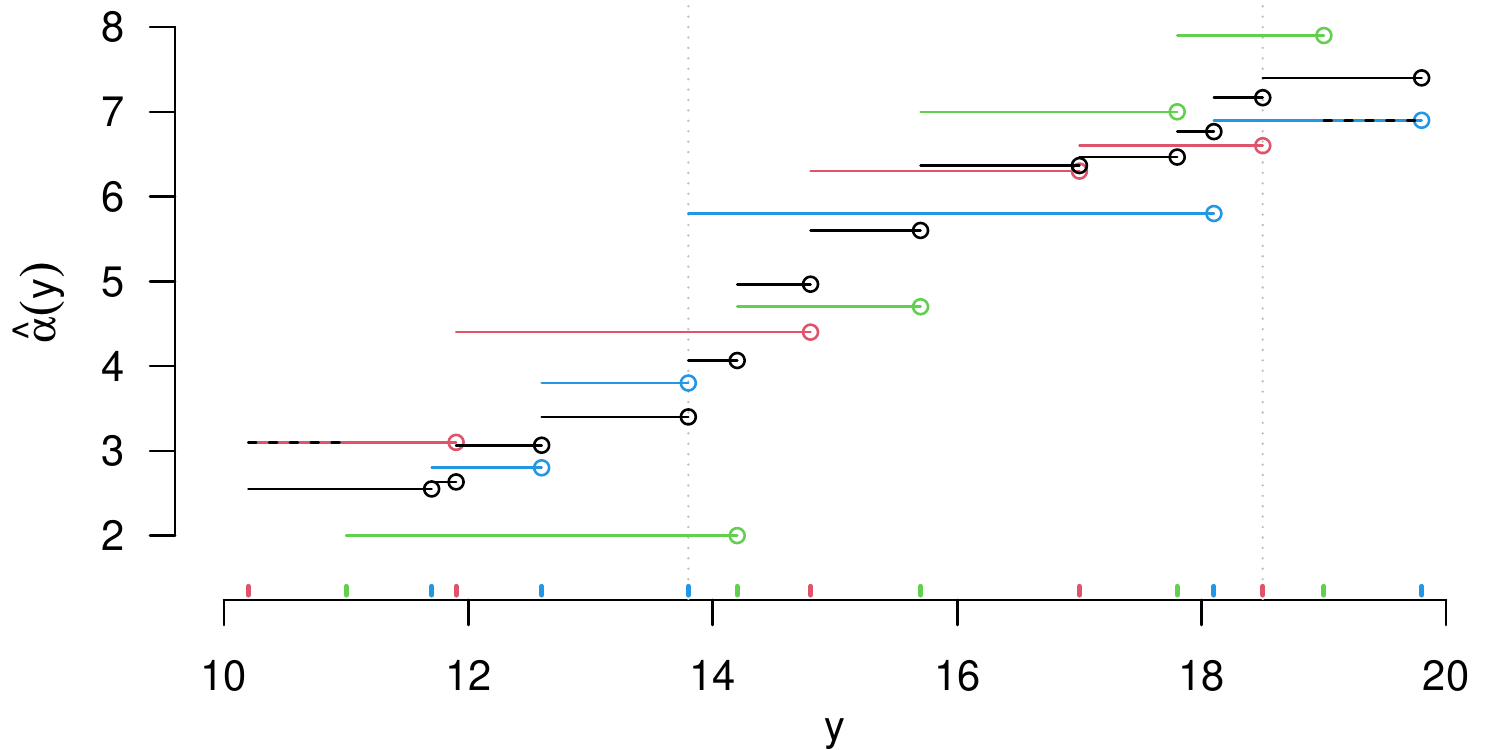}
  \end{center}
  \caption{Illustration of alpha estimation in the divide-and-combine
    approach.  Consider a hypothetical dataset with $15$ observations,
    divided into three subsets.  Each subset has $5$ observations and $4$
    alpha estimates; they are shown in red, green, and blue for subsets 1, 2,
    and 3, respectively.  The goal is to compute the $14$ alpha estimates in
    the combine step.  The $K$-vector for the outcome $y=13.8$ (indicated by
    the first vertical dotted line) is $(2,1,3)$ as it corresponds to the 2nd
    interval in subset 1, 1st in subset 2, and 3rd in subset 3.  The
    $K$-vector for $y=18.5$ (the second dotted line) is $(0,4,4)$ as it has
    no contribution from subset 1 and corresponds to the 4th interval in
    subsets 2 and 3.  The black horizontal lines are the alpha estimates in
    the combine step, and the dashed black lines at the ends are the values
    before imposing the monotonicity constraint.}
  \label{fig:alphacombine}
\end{figure}

After fitting individual CPMs to subsets, we compute the estimates of
$\bm\alpha$ (an $(M-1)$-vector) and $\bm\beta$ (a $p$-vector).  The estimates
of the beta parameters are straightforward to compute as they are the average
of the corresponding beta estimates from the subsets.  However for a
parameter in $\bm\alpha$, due to random partitioning, the corresponding alpha
estimates in the subsets can be at different index positions; we thus define
a $K$-vector for each parameter to indicate which estimate from the
individual models will be used in the computation.  Figure
\ref{fig:alphacombine} illustrates the definition.  For example, a $K$-vector
of $(3,6,\ldots,1)$ indicates that the corresponding estimate is an average
of $c_3^{(1)}, c_6^{(2)},\cdots,c_1^{(K)}$.  The first and last several
parameters in $\bm\alpha$ will not have estimates available from all the
subsets; in the $K$-vector, we use $0$ to indicate that the corresponding
model is not contributing.  There are at most $K-1$ such alpha parameters at
each end.  $K$-vectors can also be defined for the beta parameters:
$(m_1,\cdots,m_K)$ for $\beta_1$, $(m_1+1,\cdots,m_K+1)$ for $\beta_2$, etc.

Once we have defined all the $K$-vectors, we can compute the final parameter
estimates.  Given a $K$-vector $a=(a_1,a_2,\cdots,a_K)$, the corresponding
parameter estimate is an average,
\begin{equation*}
  \textstyle\left. \sum_k c^{(k)}_{a_k}I(a_k>0) \middle / \sum_kI(a_k>0). \right.
\end{equation*}
When $y\in [y_{(K)}, y_{(N-K)}]$, the average is always over $K$ values.
Since $\widehat{\bm\alpha}^{(k)}$ is non-decreasing for every $k$, the final
alpha estimates over $y\in [y_{(K)}, y_{(N-K)}]$ are also non-decreasing.
This monotonicity is not guaranteed at the ends.  To ensure monotonicity at
the ends, we apply the following constraints sequentially: Backwardly for
$i=K-1, \cdots,1$, if $\alpha_{i}>\alpha_{i+1}$ then set
$\alpha_{i}=\alpha_{i+1}$; forwardly for $i=M-K+1,\cdots, M-1$, if
$\alpha_{i}<\alpha_{i-1}$ then set $\alpha_{i}=\alpha_{i-1}$.

The estimates of $(\bm \alpha, \bm \beta)$ using this divide and combine approach for a fixed $K$ are consistent and asymptotically normal under the same conditions that are required for consistency and asymptotic normality of the standard CPM estimators. In short, under these conditions, each of the estimates in the $K$ separate analyses is consistent and asymptotically normal, and as the analyses are independent, their average is also consistent and asymptotically normal. %Note that with fixed lower and upper bounds, $L$ and $U$ respectively, and very large sample sizes, it is quite likely that a non-zero proportion of subjects in each of $K$ splits will be below $L$ or above $U$, which are common across all $K$ divisions.  Hence, the issues described above for estimating $\alpha$ in the tails, may not arise if these extreme values are lumped into categories denoting the smallest and largest values.

Because of the independence between the $K$ separate analyses, the variance-covariance matrix for
$(\widehat{\bm\alpha}, \widehat{\bm\beta})$ can be computed easily with
$K$-vectors.  Let $V^{(k)}$ be the variance-covariance matrix of
$(\widehat{\bm\alpha}^{(k)}, \widehat{\bm\beta}^{(k)})$, and $v_{i,j}^{(k)}$
be its element at $(i,j)$.  The variance of a parameter in
$(\widehat{\bm\alpha}, \widehat{\bm\beta})$ with $K$-vector
$a=(a_1,a_2,\cdots,a_K)$ is estimated as
\begin{equation*}
  \textstyle\left. 
    \sum_k v^{(k)}_{a_k,a_k} I(a_k>0) \middle/
    \left\{ \sum_kI(a_k>0) \right\}^2. \right.
\end{equation*}
The covariance between a parameter with $K$-vector $a=(a_1,a_2,\cdots,a_K)$
and another with $K$-vector $b=(b_1,b_2,\cdots,b_K)$ is estimated as
\begin{equation*}
  \textstyle\left. 
    \sum_k v^{(k)}_{a_k,b_k} I(a_k>0)I(b_k>0) \middle/
    \sum_kI(a_k>0)\cdot\sum_kI(b_k>0). \right.
\end{equation*}
%\begin{equation*}
%  (v^{(1)}_{a_1,b_1} +\cdots+ v^{(K)}_{a_K,b_K}) / (s_as_b), \text{ where }
%  s_a=\sum_kI(a_k>0) \text{ and } s_b=\sum_kI(b_k>0).
%\end{equation*}
%If $a_k=0$ or $b_k=0$, the corresponding term in the numerator is skipped.
We note that the variance estimates for the alpha parameters at the ends may
be slightly overestimated because the monotonicity constraints would likely
reduce the variation of the final estimates.  The whole variance-covariance
matrix has dimensions $(M-1+p) \times (M-1+p)$, which can be very large.  For
example, when $M=100{,}000$ and $p=100$, the matrix has more than 10 billion
numbers.  This is a challenge in both computation and storage.  In our
implementation, we calculate the diagonal values (i.e., the variances) for
the alpha portion and the full $p\times p$ variance-covariance matrix for the
beta portion.  One consequence of this choice is that the conditional mean or
median given covariate values would not have an estimated standard error as
they require covariances in the computation.

\subsection{Equal-quantile binning}

In equal-quantile binning, we first group the outcomes into equal-quantile
bins and then assign a new outcome value to the observations in each bin.
Specifically, we define a new outcome variable, $Y_b$, which for each bin
$B$, takes the value $\text{median}\{y:y\in B\}$ for all the observations in
the bin.  When there are not many ties in the original outcome, the number of
distinct values in $Y_b$ is often the number of bins, $M_b$, which can be
predetermined.  We then fit a CPM of $Y_b$ on the predictor variables.

To ensure that the $N$ observations are divided to $M_b$ bins as evenly as
possible, we implement the following algorithm: Express $N$ as
\begin{equation*}
  N=M_bq+r=(M_b-r)q+r(q+1),
\end{equation*}
where $q=\lfloor NM_b^{-1} \rfloor$ is the floor of $NM_b^{-1}$ (i.e., the
quotient) and $r$ is the remainder (i.e., $r=N \bmod M_b$); $r=0$ when $N$ is
a multiple of $M_b$.  There will be $M_b-r$ bins, each with $q$ observations,
and $r$ bins, each with $q+1$ observations.  To achieve this, we first
generate a random list of length $M_b$, in which the value $q$ occurs $M_b-r$
times and $q+1$ occurs $r$ times.  We then sort the observations by the
outcome, and group them according to the values in the random list.  For
example, if the list is $(q+1,q,\cdots)$, those with the smallest $q+1$
outcome values are in the first bin, the next set of $q$ observations are in
the second bin, etc.

One advantage of binning is that it is scale-independent with respect to the
outcome, a feature shared by the CPM itself.  Another advantage is that it
allows us to control the refinement in the new outcome variable $Y_b$ because
the number of bins can be predetermined.  Note that in the binning approach,
$\alpha(y)$ extends only to the median values of the first and last bins
instead of the ends of the original outcome.

It is well-known that with ordinal data, CPMs estimate the same beta parameter if one combines adjacent ordinal categories [REF].  Similarly, with continuous data, when one bins outcomes, one estimates the same beta parameter as if one had fit a CPM to the continuous data. Specifically, let $Y_b=g_b(Y)$, where $g_b(\cdot)$ is the binning function. From (1),  $Y_b=g_b(H(\bm\beta^T X + \epsilon)) = H_b(\bm\beta^T X + \epsilon)$ where $H_b(\cdot)=g_b(H(\cdot))$ is the resultant transformation from the latent variable scale to the binned observed outcome scale.  Because $g_b(\cdot)$ is monotonic, then $H_b(\cdot)$ is monotonic, although the inverse transformation, $\alpha_b(Y)=H_b^{-1}(Y)$, is not one-to-one because of the binned nature of the data.  The estimator for $\bm \beta$, $\widehat{\bm{\beta}}$, remains consistent and asymptotically normal. The estimator, $\widehat{\bm{\alpha}}_b$ is consistent for $\alpha_b(y)$ which is a discretized version of $\alpha(y)$. As the binning becomes finer, i.e., $M_b \rightarrow N$, $\alpha_b(y) \rightarrow \alpha(y)$. Hence, as one would expect, the accuracy of $\widehat{\bm{\alpha}}_b$ for estimating $\alpha(y)$ depends on the fineness of the binning.  With that said, with large sample sizes and continuous data, any bias in $\widehat{\bm{\alpha}}_b$ due to binning except in the tails of the distribution, is often negligible, as will be seen in Section 3.

\subsection{Rounding with refinement}

Rounding can also reduce the number of distinct outcomes.  While binning can
substantially redefine the outcomes at the extreme ends, rounding often keeps
those values nearly unchanged.  For example, suppose the largest five values
of a dataset are $\{50.3, 79.7, 130.3, 203.8, 310.7\}$, and bins are chosen
to be of size 5.  Then binning would collapse these values to their median,
$130.3$.  In contrast, rounding to integers would keep these values largely
unchanged with little loss of information.  However, rounding to a decimal
place can be a poor choice for skewed outcomes.  For example, suppose the
smallest five observations from that same dataset are
$\{0.002, 0.009, 0.019, 0.035, 0.041\}$.  Then rounding to the integer could
result in a substantial loss of information at the lower end.  For this
reason we consider two general rounding strategies: rounding to decimals and
rounding to significant digits.  We also describe a refinement step, which
allows both rounding strategies to approximately result in a target number of
distinct values.

Let $s$ be an integer, and $\lfloor a\rceil_s$ be the result of rounding a
number $a$ to decimal place $s$ (when $s>0$), to an integer (when $s=0$), or
to $10^{-s}$ (when $s<0$).  For example, $\lfloor 12.34\rceil_{1}=12.3$,
$\lfloor 12.34\rceil_{0}=12$, and $\lfloor 12.34\rceil_{-1}=10$.  When $s=0$,
we omit the subscript and write $\lfloor a\rceil$.  Let
$\lfloor a\rceil^{(s)}$ ($s>0$) be the result of rounding $a$ to $s$
significant digits.  Let $p(a)=\lfloor\log_{10}a \rfloor$ be the place of the
first significant digit of $a$.  Then
$\lfloor a\rceil^{(s)} =\lfloor a\rceil_{s-1-p(a)}$.  For example,
$\lfloor 12.34\rceil^{(2)} =\lfloor 12.34\rceil_{0}=12$.

We now refine these two types of rounding.  When we round $a$ to an integer,
$\lfloor a\rceil$ is the integer that is closest to $a$.  We could refine
this by rounding $a$ to the closest multiple of $0.5$, which is effectively
$\lfloor 2a\rceil/2$.  Similarly, $\lfloor 3a\rceil/3$ is to round $a$ to the
closest multiple of $1/3$.  In general, for any real number $t\in[1,10]$,
$\lfloor ta\rceil/t$ is to round $a$ to the closest multiple of $1/t$.  As
$t$ increases from $1$ to $10$, more refined rounding is achieved; when
$t=10$, we reach $\lfloor a\rceil_1$.  For any integer $s$ and any number
$t\in[1,10]$, we define
\begin{equation*}
  \lfloor a\rceil_{s,t} =\lfloor ta\rceil_{s}/t
\end{equation*}
as the value after rounding $a$ to place $s$ at refinement level $t$.  When
$t=1$, $\lfloor a\rceil_{s,1} =\lfloor a\rceil_{s}$; as $t$ increases,
$\lfloor a\rceil_{s,t}$ becomes more refined and
$\lfloor a\rceil_{s,10}=\lfloor a\rceil_{s+1}$.  Similarly, for any integer
$s>0$ and any number $t\in[1,10]$, we define
\begin{equation*}
  \lfloor a\rceil^{(s,t)} =\lfloor a\rceil_{s-1-p(a),t}
\end{equation*}
as the value after rounding $a$ to $s$ significant digits at refinement level
$t$.  When $t=1$, $\lfloor a\rceil^{(s,1)} =\lfloor a\rceil^{(s)}$; as $t$
increases, $\lfloor a\rceil^{(s,t)}$ becomes more refined and
$\lfloor a\rceil^{(s,10)}=\lfloor a\rceil^{(s+1)}$.

We now describe the two rounding algorithms for a given dataset.  Let
$m_{s,t}$ be the number of distinct values after rounding the outcomes to
place $s$ at refinement level $t$.  Similarly, let $m^{(s,t)}$ be that after
rounding to $s$ significant digits at refinement level $t$.  Let
$M_r$ be a target number of distinct outcomes after rounding.  Our two
rounding algorithms are:\\
{\bf Decimal place rounding with refinement}:
\begin{enumerate}
\item[(Ia)] Identify $s$ such that $m_{s,1}\le M_r<m_{s+1,1}$;
\item[(Ib)] If $m_{s,1}=M_r$, set $t=1$.  Otherwise, search over a grid in
  $[1,10]$ to identify $t$ such that $m_{s,t}$ is closest to $M_r$, as
  measured by the smallest $\vert \log m_{s,t}-\log M_r \vert$;
\item[(Ic)] Define a new outcome variable $Y_r$ such that its value is
  $\lfloor y_i\rceil_{s,t}$ for observation $i$.
\end{enumerate}
{\bf Significant digit rounding with refinement}:
\begin{enumerate}
\item[(IIa)] Identify $s$ such that $m^{(s,1)}\le M_r<m^{(s+1,1)}$;
\item[(IIb)] If $m^{(s,1)}=M_r$, set $t=1$.  Otherwise, search over a grid in
  $[1,10]$ to identify $t$ such that $m^{(s,t)}$ is closest to $M_r$, as
  measured by the smallest $\vert \log m^{(s,t)}-\log M_r \vert$;
\item[(IIc)] Define a new outcome variable $Y_r$ such that its value is
  $\lfloor y_i\rceil^{(s,t)}$ for observation $i$.
\end{enumerate}

In our implementation of these rounding algorithms, the grid for $t$ has
increment of $0.1$, and the resulting numbers of distinct values in $Y_r$
have been always ${<}1\%$ different from the target number $M_r$.  Compared
to the binning approach, these rounding approaches do not require sorting of
the outcomes, but they are scale-dependent with respect to the outcome.

The theoretical justification for rounding is identical to that for binning because they both merge adjacent values in discretized categories.

\section{Simulations and Results}
\subsection{Simulation setup}

We simulated datasets with various sample sizes and with $p=50$ predictor
variables.  Half of the predictors are binary with success probabilities
ranging from 0.05 to 0.5, and half are continuous from normal distributions
with mean ranging from 0 to 2.4 and variance 1.  The beta coefficients for
half of the predictors were specified in the range $[-1,1]$ and those for the
other predictors were set as zero.  We then generated
$y^*=\beta^T x+\epsilon$, where $\epsilon\sim F_{\epsilon}$, and $y=H(y^*)$.
We considered six scenarios, corresponding to the combinations of three
options for $H(\cdot)$ and two options for $F_{\epsilon}$.  The three options
for $H(\cdot)$ are: (1) $y=y^*$, for which the inverse is $\alpha(y)=y$; (2)
$y=\exp(y^*)$, for which $\alpha(y)=\log(y)$; and (3) $y=\log(y^*+y^*_0)$,
where $y^*_0$ is added to ensure $y^*+y^*_0>0$ for all the simulated $y^*$;
for this transformation, $\alpha(y)=e^y-y^*_0$.  The two options of
$F_{\epsilon}$ are: (i) the logistic distribution
$F_{\epsilon}(u)=e^u/(1+e^u)$; (ii) the Gumbel distribution
$F_{\epsilon}(u)=\exp(-e^{-u})$.

We simulated datasets with sample size $N=10^4$, $4\times 10^4$ and $10^6$.
The smaller sample sizes allow us to compare the three new approaches with
the ``gold standard'' approach of fitting a single CPM to the original data.
We also compared the parameter estimates with the true parameters used in
data generation.  In the divide-and-combine approach, the data were
partitioned into $K=10$ subsets when $N=10^4$, $40$ when $N=4\times10^4$, and
$100$ when $N=10^6$.  In the binning approach, the outcomes were grouped into
$M_b=1{,}000$ bins when $N=10^4$, and $10{,}000$ bins otherwise.  In the
rounding approach, the outcomes were rounded with a target number of
$M_r=1{,}000$ when $N=10^4$, and $10{,}000$ otherwise.  These settings are
displayed in Table \ref{tab:simulationsetup}.

\begin{table}
  \caption{Evaluation set-up of simulated data}
  \begin{center}
    \begin{tabular}{ccccccc}
      \hline
      $N$ & $p$ & Truth & Whole data & Divide-and-combine & Binning & Rounding \\ \hline
      $10^4$ & $50$ & \checkmark & \checkmark & \checkmark ($K=10$) & \checkmark ($M_b=10^3$) & \checkmark ($M_r\sim 10^3$) \\
      $4\times10^4$ & $50$ & \checkmark & \checkmark & \checkmark ($K=40$) & \checkmark ($M_b=10^4$) & \checkmark ($M_r\sim 10^4$) \\
      $10^6$ & $50$ & \checkmark & & \checkmark ($K=100$) & \checkmark ($M_b=10^4$) & \checkmark ($M_r\sim 10^4$) \\
      \hline
    \end{tabular}
  \end{center}
  \label{tab:simulationsetup}
\end{table}

The endpoints we evaluated include: (1) the estimates of the parameters in
$\bm\alpha$ and $\bm\beta$, and their standard errors; (2) estimates of
conditional mean and median at 10 sets of $x_0$ randomly selected from the
simulated datasets.  Specifically, let $y_{(j)}$ ($j=1,\cdots,M$) be the
$j$-th smallest distinct outcome value.  Given $X=x_0$, the cumulative
distribution function of $Y$ is estimated as
$\hat P_j =\Pr(Y\le y_{(j)} | x_0)
=F_{\epsilon}(\hat\alpha_j-\hat\beta^Tx_0)$ when $j<M$, and
$\hat P_M =\Pr(Y\le y_{(M)} | x_0)=1$ when $j=M$.  The corresponding
probability mass function is estimated as
$\hat p_j =\Pr(Y=y_{(j)} | x_0) =\hat P_j- \hat P_{j-1}$ ($j=1,\cdots,M$),
where $\hat P_0=0$.  We then computed the conditional mean,
$\sum_{j=1}^M \hat p_j y_{(j)}$;
% and a trimmed mean, $m_2 =\sum_j \hat p_j y^{(j)} / \sum_j \hat p_j$, where
% the sums are over distinct outcome values excluding those in the smallest
% $0.5\%$ and the largest $0.5\%$ cumulative probabilities.
and the conditional median as the average of the two adjacent $y_{(j)}$ and
$y_{(j+1)}$ for which $\hat P_j<0.5$ and $\hat P_{j+1}>0.5$.  The true
conditional mean and median at $X=x_0$ were empirically obtained by first
generating $10{,}000$ values of $\epsilon\sim F_{\epsilon}$ and corresponding
$H(\beta^T x_0+\epsilon)$, and then computing their mean and median.
% We also computed the standard error for $m_1$ when $N=10^4$.

\subsection{Simulation results}

\begin{figure}[!tb]
  \begin{center}
    \includegraphics[width=6in]{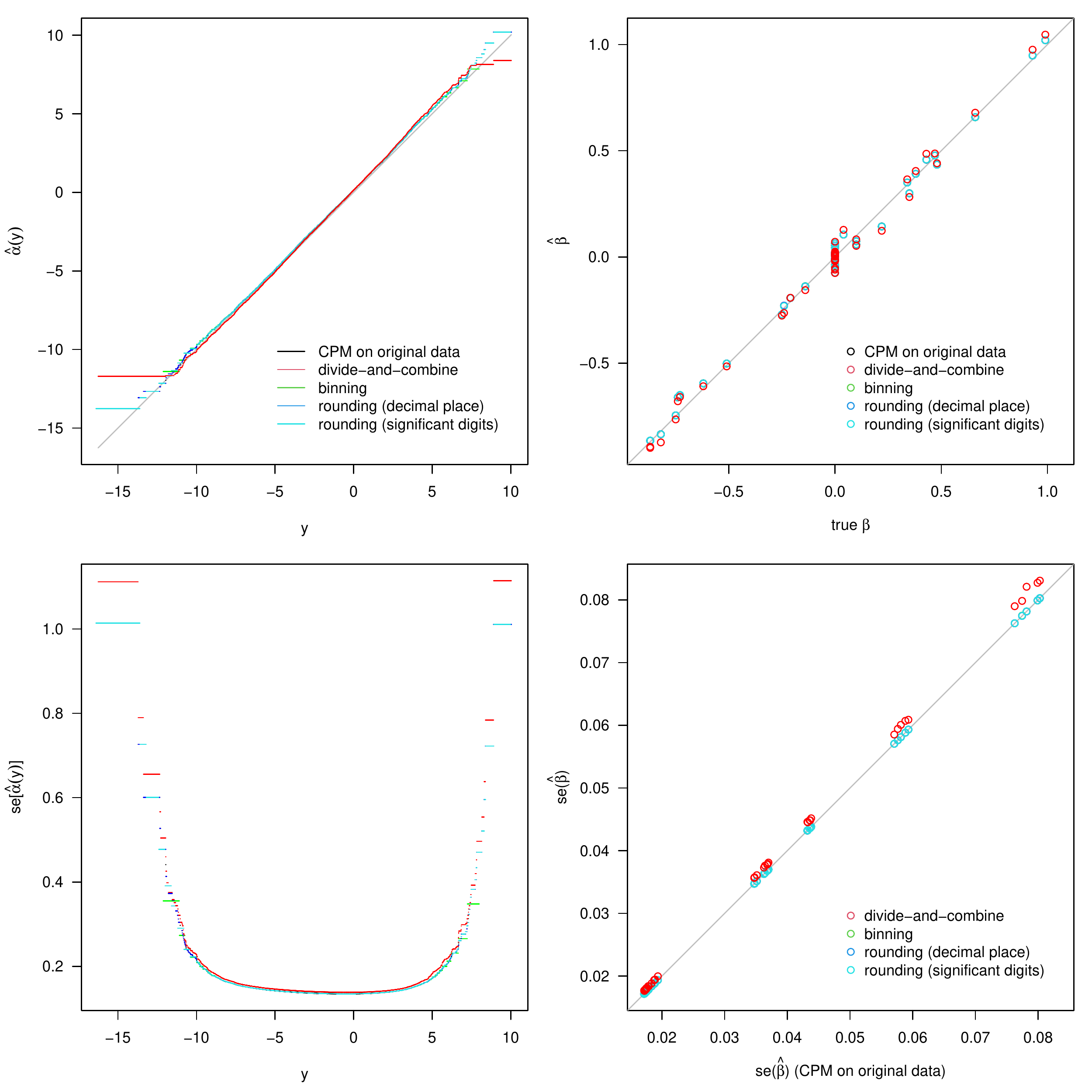}
  \end{center}
  \caption{Estimation of parameters and their standard errors under the
    scenario of logistic residual distribution, $y=H(y^*)=y^*$, $N=10^4$.
    Left panel: alpha estimates as functions of the outcome (top) and their
    standard errors (bottom).  The gray line is the truth: $\alpha(y)=y$.
    The rounding approaches and the CPM on the original data had nearly
    identical results.  Right panel: beta estimates (top) and their standard
    errors (bottom).  Gray diagonal lines $y=x$ are added for reference.  The
    binning and rounding approaches and the CPM on the original data had
    nearly identical results.}
  \label{fig:alphabeta10K}
\end{figure}

Figure \ref{fig:alphabeta10K} shows the parameter estimates and their
standard errors from the three new approaches for the simulation scenario
with logistic residual distribution, identity transformation $y=H(y^*)=y^*$,
and $N=10^4$.  The results of the ``gold standard'' approach of fitting a CPM
with the whole dataset are also shown.

The $\hat\alpha(y)$ from all the approaches agreed very well with the true
$\alpha(y)$ over a wide range of $y$ from ${<}0.5$ percentile to ${>}99.5$
percentile.  The standard error for $\hat\alpha(y)$ was also relatively low
in this range due to the high data density.  There was some departure of
$\hat\alpha(y)$ from the truth at the ends, with relatively high standard
errors, probably due to a lack of information as a result of data sparcity.

In the divide-and-combine approach, $\hat\alpha(y)$ started to nearly flatten
out at $y_{(K)}$ and $y_{(N-K)}$.  This approach also yielded slightly higher
standard errors than the other approaches.  In the binning approach, by
definition, $\hat\alpha(y)$ extends only to the median values of the first
and last bins.  The rounding to decimal place approach had nearly identical
estimates of the alpha parameters and their standard errors as the ``gold
standard'' approach.  This is because the outcome in this simulation scenario
had a relatively symmetric distribution, for which rounding to a decimal
place tends to have little effect at the ends where the values are relatively
large.  In comparison, rounding to significant digits tends to have a
slightly larger effect at the ends.

The estimates of the beta parameters were also quite accurate for all
approaches.  The binning and rounding approaches and the ``gold standard''
approach had nearly identical estimates of the beta parameters and their
standard errors.  In comparison, the divide-and-combine approach yielded
slightly different parameter estimates and slightly higher standard errors.

The results for the other transformations and for the Gumbel residual
distribution showed similar patterns (Supplemental Material).  Note that for
very skewed outcomes, the rounding to significant digits approach had nearly
identical estimates of all the parameters and their standard errors as the
``gold standard'' approach, while rounding to a certain decimal place rounded
many values at the lower end to a single value, which is undesirable.

\begin{figure}[!tb]
  \begin{center}
    \includegraphics[width=6in]{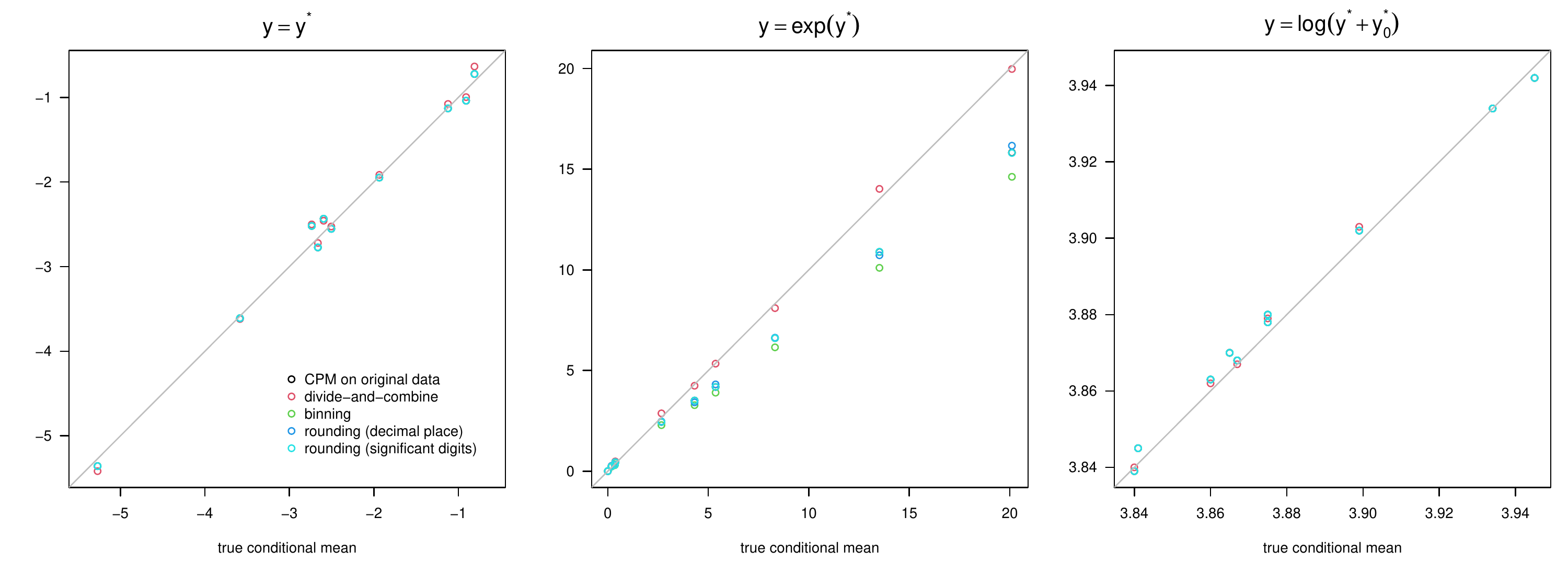}
    \includegraphics[width=6in]{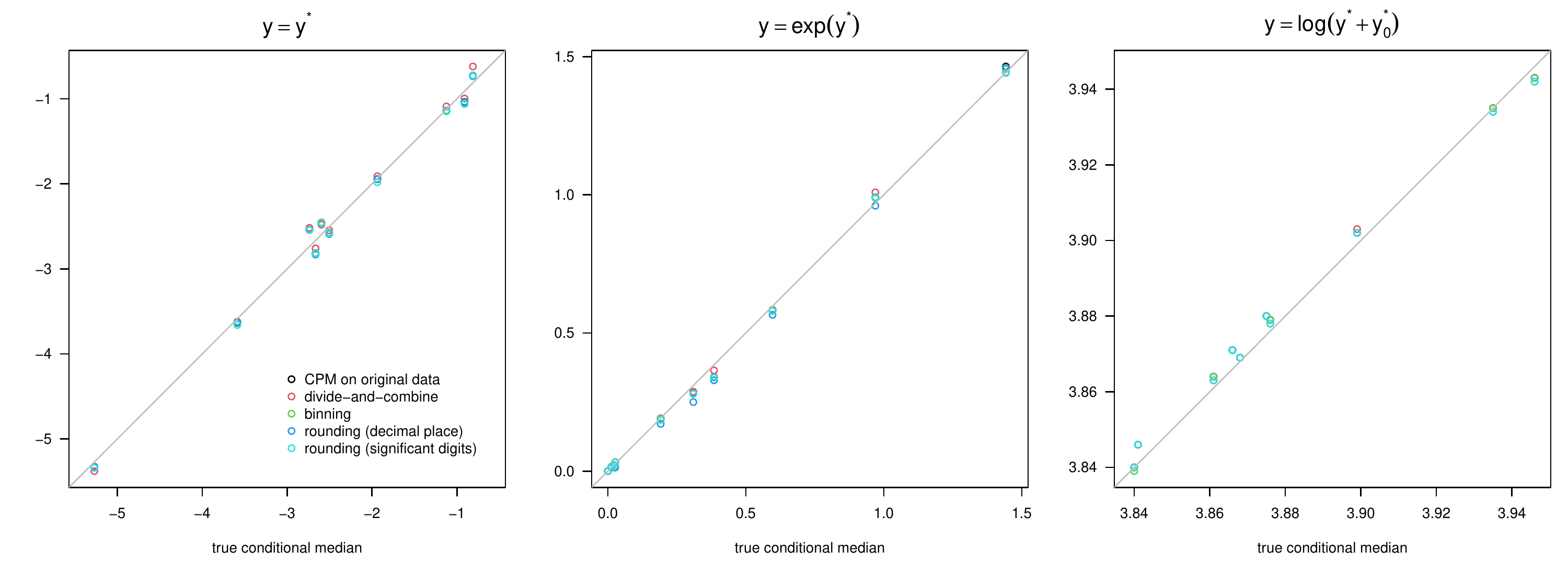}
  \end{center}
  \caption{Estimation of conditional mean (top) and conditional median
    (bottom) for 10 sets of $x$.  The columns are for the three simulation
    scenarios with outcome transformations $y=y^*$, $y=\exp(y^*)$, and
    $y=\log(y^*+y^*_0)$, respectively, all with logistic residual
    distribution and $N=10^4$.}
  \label{fig:condmean}
\end{figure}

Figure \ref{fig:condmean} shows the estimated conditional mean and median for
10 sets of $x$ randomly selected from the simulated datasets.  When the
conditional distribution is relatively symmetric, as in the scenarios with
$y=y^*$ and $y=\log(y^*+y^*_0)$, the estimation of conditional mean performed
well.  When the conditional distribution is very skewed, as in the scenario
with $y=\exp(y^*)$, the estimation of conditional mean performed poorly.  We
repeated the simulations under the latter scenario with multiple replicates
and found a high variation in the results, presumable due to high variation
in the alpha estimates at the high end of the outcomes.  The pattern we
observed seems to be overestimation from the divide-and-combine approach and
underestimation from the other approaches including the ``gold standard'' CPM
on the original data.  In contrast, the conditional medians performed quite
well for all the approaches in all the scenarios.  The results for the Gumbel
residual distribution have similar patterns (Supplemental Material).

\begin{figure}[!tb]
  \begin{center}
    \includegraphics[width=6in]{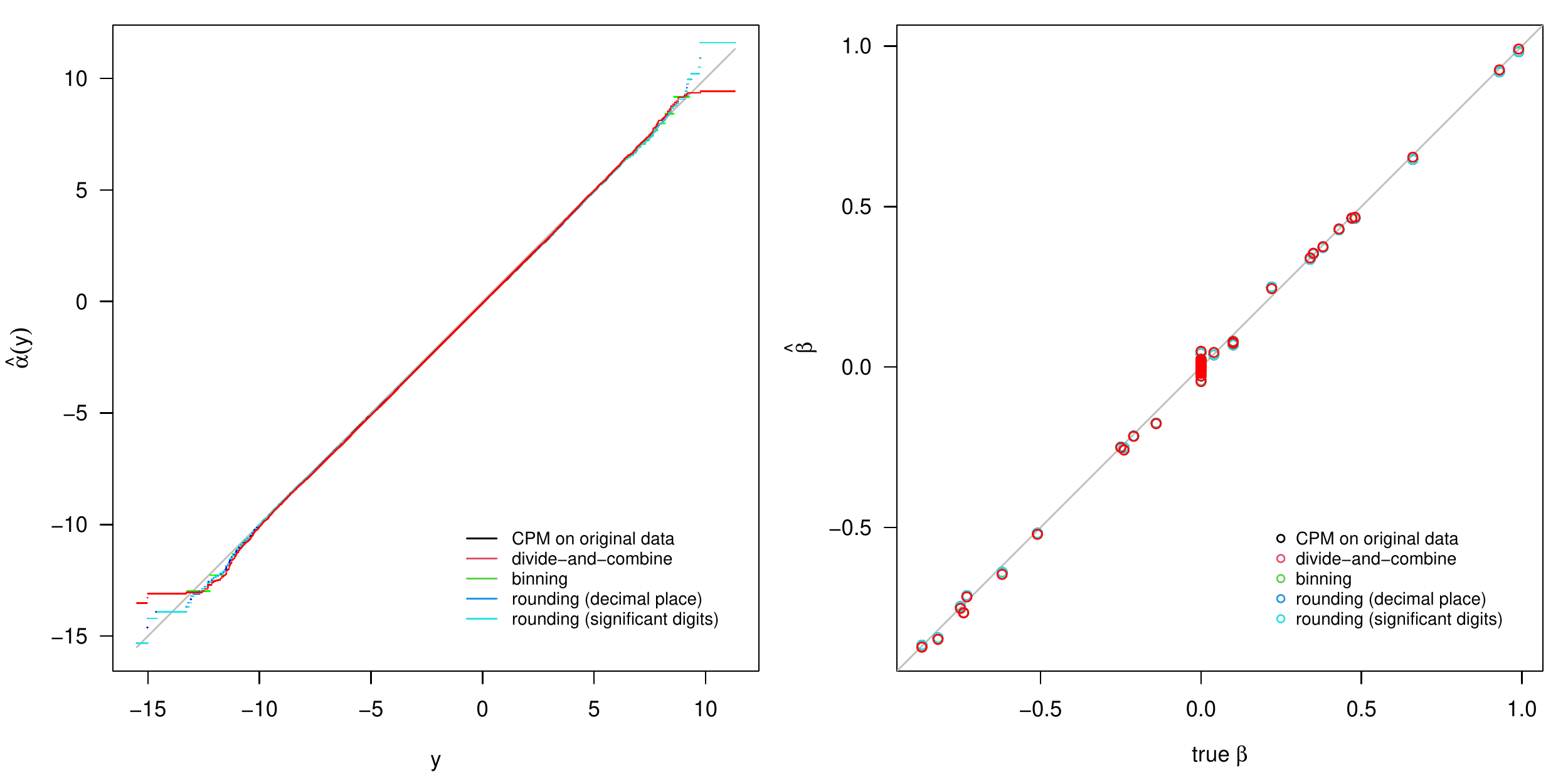}
    \includegraphics[width=6in]{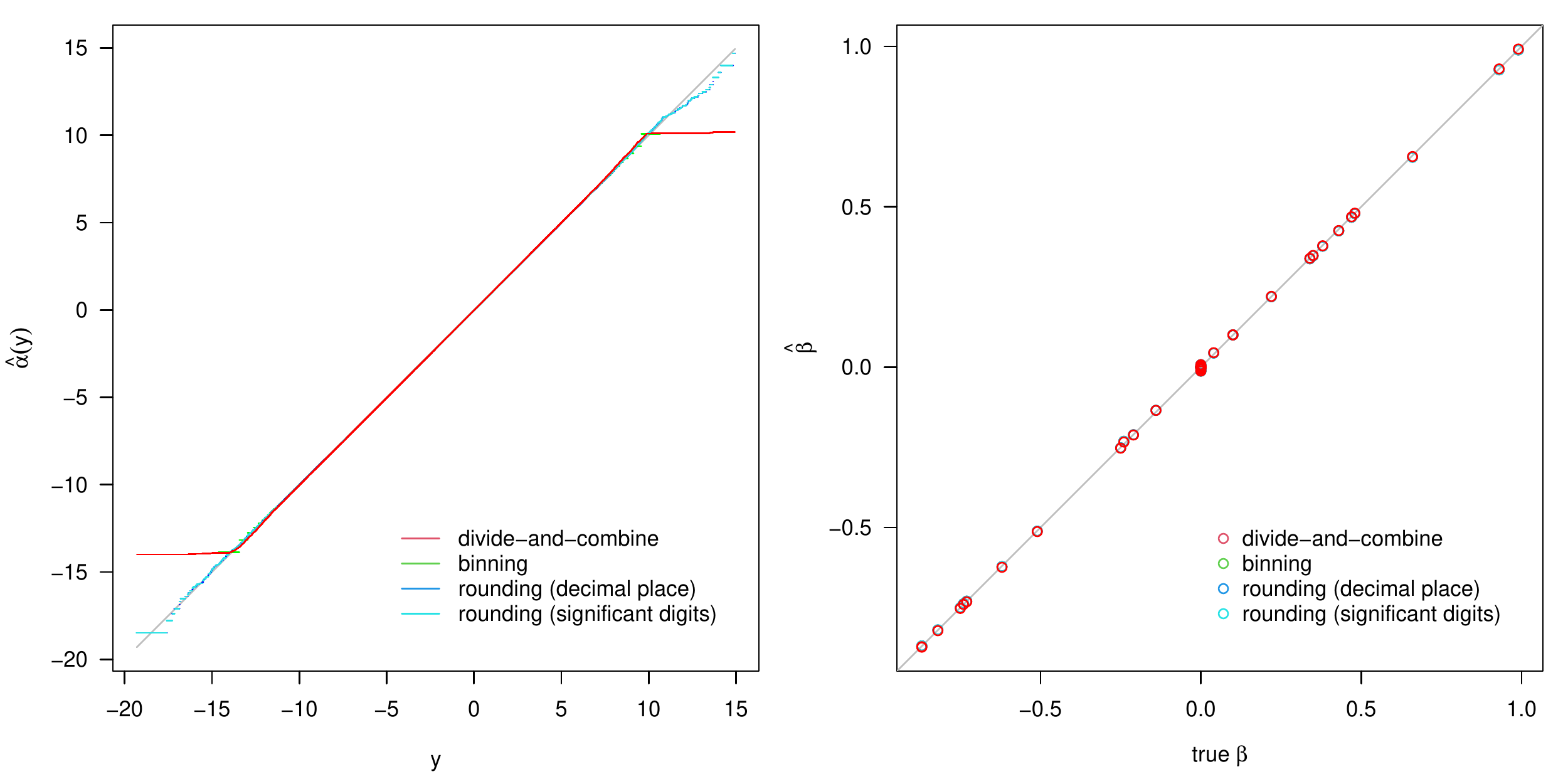}
  \end{center}
  \caption{Estimation of parameters for large sample sizes under the scenario
    of logistic residual distribution and $y=H(y^*)=y^*$.  Top row:
    $N=4\times10^4$.  Bottom row: $N=10^6$.  It was impossible to fit a
    single CPM on the original data with $N=10^6$.  Left: alpha estimates as
    functions of the outcome.  The gray line is the truth: $\alpha(y)=y$.
    Right: beta estimates.  Gray diagonal lines $y=x$ are added for
    reference.}
  \label{fig:40K1M}
\end{figure}

Figure \ref{fig:40K1M} shows the estimates of the parameters for sample sizes
$4\times10^4$ and $10^6$ with logistic residual distribution and
$y=H(y^*)=y^*$.  In comparison to the results for $N=10^4$ in Figure
\ref{fig:alphabeta10K}, a larger sample size clearly led to a smaller
difference between the estimates and the truth, illustrating that the
approaches yield consistent parameter estimates.  The standard error
estimates also became smaller as the sample size increased (Supplemental
Material).

The divide-and-combine approach introduces variation due to random
partitioning.  However, the variation due to random partitioning is minor
relative to that from random sampling (see details in Supplemental Material).

\subsection{Time and peak memory usage}

To help determine the number of subsets in the divide-and-combine approach
and the target number of distinct values in the binning and rounding
approaches, we evaluate the running time and peak memory usage when fitting a
single CPM with respect to the sample size, the number of distinct outcome
values, and the number of predictors.  We generated datasets with various
values of $N$ (sample size: $N=5000, 10000, 20000, 30000, 40000$), $M$
(number of distinct outcome values: $M=1000, 5000, 10000, 20000, 40000$), and
$p$ (number of predictors: $p=10,25,50,100$).  All variables were continuous.
When $M<N$, equal-quantile binning was used to achieve the desired $M$.
There are 18 combinations of $N$ and $M$ for each value of $p$, and there are
a total of 72 datasets.  We fit \verb|orm()| to each dataset and recorded the
running time and peak memory usage.

\begin{figure}[!t]
  \begin{center}
    \includegraphics[width=6in]{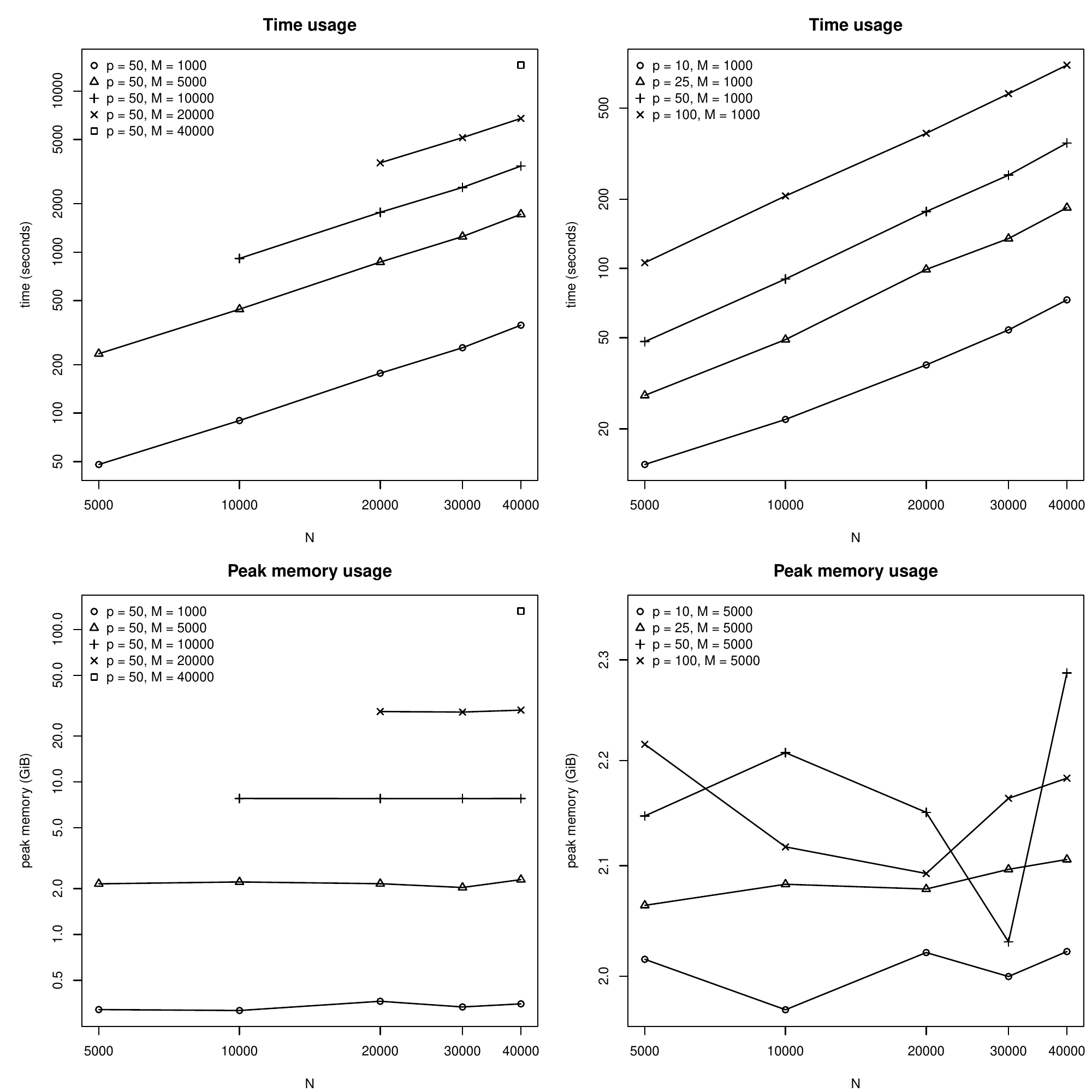}
  \end{center}
  \caption{Time and peak memory usage for fitting orm() to datasets with
    various $N$, $M$, and $p$.  The axes are drawn on the log scale.}
  \label{fig:timememoryusage}
\end{figure}

\begin{figure}[!t]
  \begin{center}
    \includegraphics[width=6in]{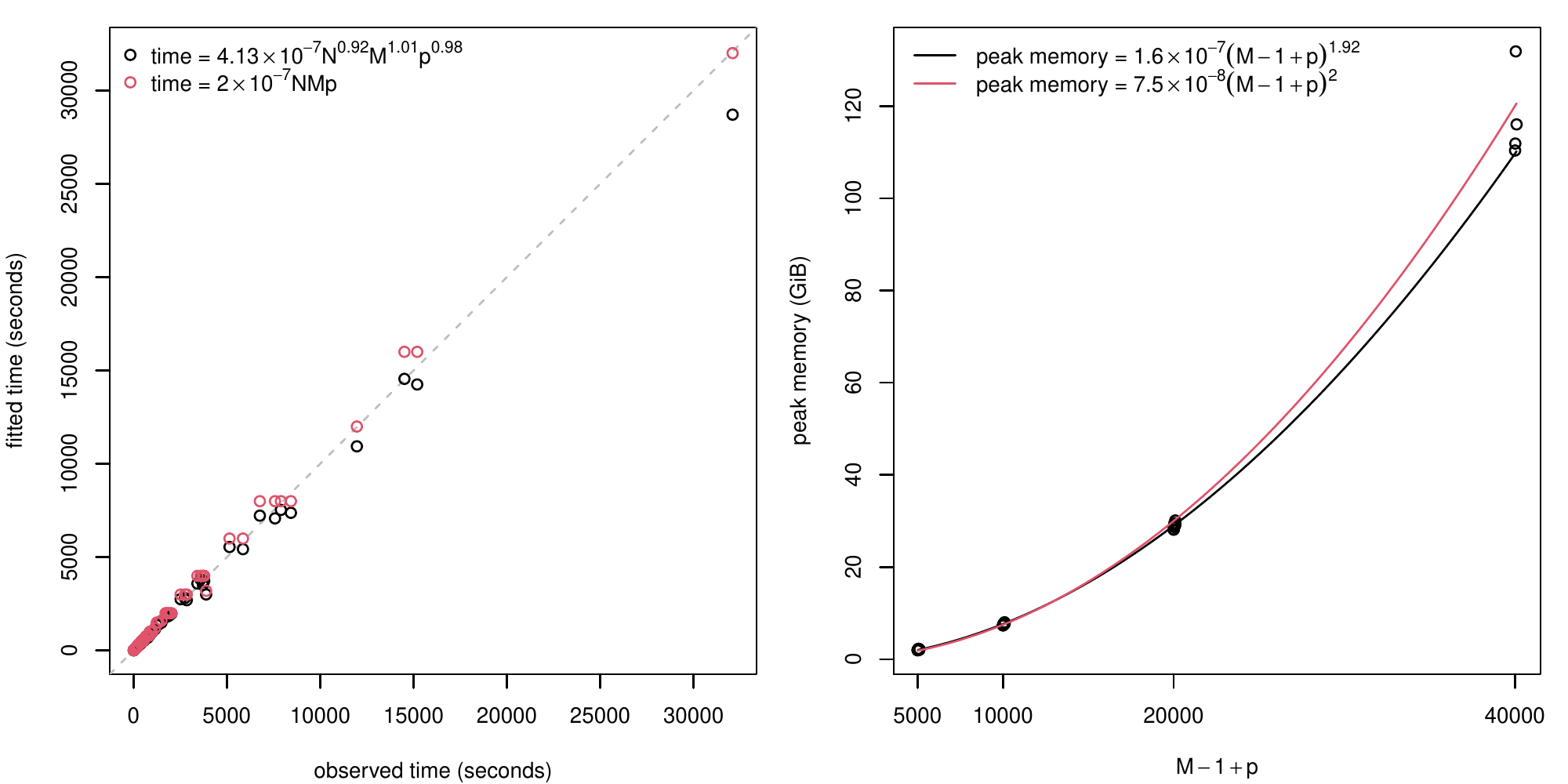}
  \end{center}
  \caption{Fitted models for time and peak memory usage.  Left: Observed
    running time vs. fitted values from two models.  Right: Peak memory usage
    vs. $M-1+p$, and curves from two models.}
  \label{fig:timememorymodels}
\end{figure}

The top row in Figure \ref{fig:timememoryusage} shows the time usage with
respect to changes in $N$ and $M$ when $p=50$ (left), and with respect to
changes in $N$ and $p$ when $M=1000$ (right).  The patterns are similar for
other values of $p$ and $M$.  The results indicate a log-log linear
relationship of time on $N$, $M$, and $p$.  We therefore fit a log-log linear
model, which is multiplicative on the time scale:
$\text{time} = 4.13\times 10^{-7}\cdot N^{0.92}M^{1.01}p^{0.98}$; this model
had $R^2=0.998$.  Thus time is approximately proportional to $NMp$; we found
that $\text{time} = 2\times 10^{-7}\cdot NMp$ seems to fit well to our
results.  The left panel of Figure \ref{fig:timememorymodels} shows the
comparison of the observed running time and the fitted values from these two
models.  Note that the constant factor $2\times 10^{-7}$ depends on our
server configuration and may vary greatly across hardware configurations.

The bottom row in Figure \ref{fig:timememoryusage} shows the peak memory
usage under the same settings as in the top row; again, the results for other
values of $p$ and $M$ have similar patterns.  The results suggest that peak
memory usage is influenced heavily by $M$, to a much less degree by $p$, and
almost ignorably by $N$.
%We fit a log-log linear model of memory on $M-1+p$:
%$\text{memory} = 7.71\times 10^{-6}\cdot (M-1+p)^{1.51}$, which had
%$R^2=0.976$.
An examination of the log-log plot of peak memory usage versus $M-1+p$ (in
Supplemental Material) suggested that the peak memory usages for datasets
with $M\ge5000$ fall along a line while those for $M=1000$ are far above that
line, presumably due to overhead operations that can dominate memory usage
for a relatively small $M$.  We thus fit a log-log linear model to the
results with $M\ge5000$, which led to:
$\text{memory} = 1.60\times 10^{-7}\cdot (M-1+p)^{1.92}$, with $R^2=0.999$.
This model suggests that peak memory usage is approximately proportional to
$(M-1+p)^2$, at least for $M\ge5000$.  This makes sense as the memory usage
is probably devoted mostly to the Hessian matrix, which has dimensions
$(M-1+p)\times(M-1+p)$.  For our data,
$\text{memory} = 7.5\times 10^{-8}\cdot (M-1+p)^2$ seems to fit well.  The
right panel of Figure \ref{fig:timememorymodels} shows the peak memory usage
versus $M-1+p$, and the curves of these two models.

% Again, the constant factor $7.5\times 10^{-8}$ depends on the server
% configuration and may vary greatly across hardware configurations.

The results that the running time of a single CPM is approximately
proportional to $NMp$ and that the peak memory usage is approximately
proportional to $(M-1+p)^2$ is helpful when determining the number of subsets
in the divide-and-combine approach and the target number of distinct values
in the binning and rounding approaches.  In the divide-and-combine approach,
a subset has much smaller $N$ and $M$ than those in the original data, and
the individual jobs take much less time and memory to run.  This also allows
simultaneous model fitting for multiple subsets, further speeding up the
process.  In the binning and rounding approaches, $M_b$ and $M_r$ are
typically much smaller than $M$ in the original data.  As a result, these
approaches also take much less time and memory, making it feasible to fit a
CPM even with a large sample size.

% We also recorded time for the aggregation step in the divide-and-combine
% approach.

\section{A Data Example}

We applied our approaches to a publicly available dataset, in which an
algorithm for multiplying two matrices of dimensions $2048\times2048$ was
evaluated for its running time on a parameterizable SGEMM GPU kernel
(Nugteren and Codreanu 2015; Ballester-Ripoll et al. 2017).  The algorithm
has 14 parameters, and $241{,}600$ parameter combinations were feasible due
to various kernel constraints.  For each combination, 4 runs were performed
and the running time was recorded in milliseconds.  For this dataset,
$N=966{,}400$ and $p=14$, and there are $M=106{,}799$ distinct outcome
values.  More details can be found at the University of California Irvine
Machine Learning Repository
(\verb|http://archive.ics.uci.edu/ml/datasets/SGEMM+GPU+kernel+performance|).

The outcome variable, the algorithm running time, is very skewed, with
skewness 3.93 and a range from 13.25 to 3397.08 milliseconds.  One could
apply a transformation (e.g., logarithm) on the outcome and then fit a linear
model, not knowing if the transformation would be a good choice.  The CPM
instead allows us to obtain a suitable transformation empirically.

\begin{table}
  \caption{Outcome categorization after rounding in the SGEMM dataset}
  \begin{center}
    \begin{tabular}{lrrrrrrrr}
      \hline
      range & \# obs & \multicolumn{6}{c}{\# distinct values} \\ \cline{3-8}
            & &$y_i$ & $\lfloor y_i\rceil$ & $\lfloor y_i\rceil_{1}$
                     & $\lfloor y_i\rceil^{(3)}$ & $\lfloor y_i\rceil^{(4)}$
                     & $\lfloor y_i\rceil^{(3,5.2)}$ \\
      $[10, 100)$    & 563831 &  8453 &   88 &   862 & 862 & 8453 & 4424\\
      $[100, 1000)$  & 368366 & 70217 &  900 &  8855 & 900 & 8855 & 4635\\
      ${>}1000$      &  34203 & 28129 & 1875 & 12356 & 207 & 1875 & 1009\\
      Total          & 966400 &106799 & 2863 & 22073 &1969 &19183 &10068\\
      \hline
    \end{tabular}
  \end{center}
  \label{tab:roundingSGEMM}
\end{table}

We fit CPMs of the outcome on the 14 predictor variables with the three
methods described in Section 2.  In the divide-and-combine approach the data
were divided into $K=48$ subsets of size $20{,}133$ or $20{,}134$.  In
binning and rounding, the target number of distinct outcome values was set as
$10{,}000$, and the final numbers of distinct outcome values were $9{,}067$
and $10{,}068$, respectively.  For rounding, we chose to round to significant
digits.  The rationale behind this decision can be seen in Table
\ref{tab:roundingSGEMM}, where we divide the outcomes into three regions:
$[10, 100)$, $[100, 1000)$, and those ${>}1000$.  The 2nd and 3rd columns
display the number of observations and the number of distinct outcome values
in these regions.  For this dataset, rounding to a decimal place (columns
4--5) would result in very few distinct values representing the $563{,}831$
outcomes in $[10,100)$, but a lot more distinct values representing the
$34{,}203$ outcomes that are ${>}1000$.  In contrast, rounding to a certain
number of significant digits (columns 6--8) gives more balanced
categorizations of the outcome.  The selected rounding scheme was rounding to
$3$ significant digits at refinement level $5.2$ (column 8), resulted in a
fairly good balance and $M_r \approx 10{,}000$.  The divide-and-combine
approach took 2 hours 37 minutes on our server while the binning and rounding
approaches each took 7 hours 12 minutes.

\begin{figure}[!t]
  \begin{center}
    \includegraphics[width=6in]{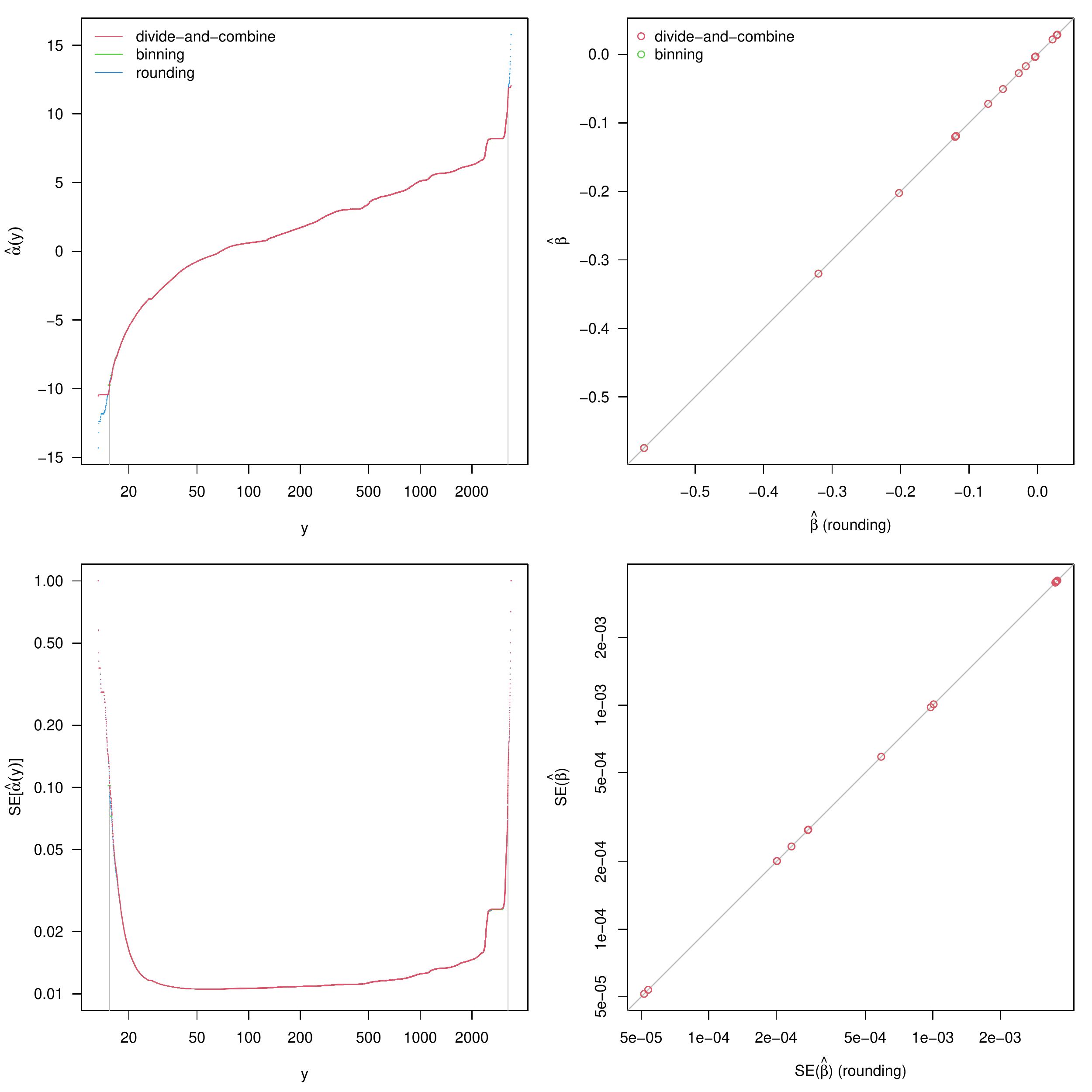}
  \end{center}
  \caption{Results of the three approaches on the SGEMM dataset.  Left:
    Estimates of $\alpha(y)$ (top) and their standard errors (bottom).  The
    $0.01$ and $99.99$ percentiles of the outcome are marked by the gray
    vertical lines.  Right: Estimates of beta parameters (top) and their
    standard errors (bottom).  Gray diagonal lines $y=x$ are added for
    reference.}
  \label{fig:dataexample}
\end{figure}

The results are summarized in Figure \ref{fig:dataexample}.  The top left
plot shows the alpha estimates as functions of the outcome.  The outcome
range was $[13.25, 3397.08]$ in the original data, $[15.15, 3268.00]$ after
binning, and $[13.25, 3396.15]$ after rounding.  The three alpha estimates
agreed remarkably well with each other from the $0.01$ percentile to the
$99.99$ percentile of the outcome $[15.41,3248.05]$.  Note that
$\hat\alpha(y)$ is an estimated transformation of the outcome such that the
transformed value would relate to the predictors linearly.  Here all three
$\hat\alpha(y)$ are very different from a log transformation, which would be
a straight line on this plot as the x-axis is on the log scale.  This
indicates that the true transformation must be very different from logarithm,
which would often be used to transform a skewed outcome in a traditional
analysis.  The bottom left plot shows the standard errors of the alpha
estimates, which are also very similar from $0.01$ to $99.99$ percentiles.
The top right plot of Figure 3 shows comparisons of the beta estimates from
the three methods, with the results of rounding on the x-axis and those of
the other two methods on the y-axis.  Again, the three methods' beta
estimates and standard errors (bottom right plot) agree remarkably well.

% All 14 predictors were statistically significant; the highest nominal
% $p$-value was $8.82\times10^{-47}$.

\section{Discussion}

Cumulative probability models (CPMs) are a robust alternative to linear
models.  However they are not feasible for very large datasets as the running
time and memory usage increase with the sample size, the number of distinct
outcomes, and the number of predictors.  In this paper, we addressed this
problem with three approaches.  The divide-and-combine approach focuses on
reducing the sample size of individual CPMs, and the binning and rounding
approaches focus on reducing the number of distinct outcome values.  With
computer simulations, we showed that these approaches perform quite well,
with estimates of parameters and their standard errors very similar to those
from a single CPM on the whole dataset (when the latter is feasible), and
with consistent parameter estimates.  The rounding approach has two
algorithms, rounding to a decimal place (for not-too-skewed outcomes) and
rounding to significant digits (for skewed outcomes).  Both algorithms have a
refinement step to achieve the desired number of distinct outcomes.  All
three approaches yielded comparable results when applied to a large dataset
with nearly one million observations.

We also studied the running time and peak memory usage in relation to the
sample size $N$, the number of distinct outcomes $M$, and the number of
predictors $p$.  We showed that the running time is approximately
proportional to $NMp$, and that the peak memory usage is approximately
proportional to $(M-1+p)^2$.  These results can help plan the analysis by
determining the number of subsets in the divide-and-combine approach and the
number of target distinct outcomes in the binning and rounding approaches.
There is a trade-off between speed and accuracy.  Therefore, for the
divide-and-combine approach, we recommend as few subsets as allowed by
computer resources, and for the binning and rounding approaches, we recommend
as large target number of distinct outcomes as allowed by computer resources.

There are some limitations in the divide-and-combine approach.  When the
sample size is very large, computation of the covariances in the
variance-covariance matrix will be infeasible as it requires a large amount
of storage space.  In addition, when a predictor variable has a vast majority
of the observations having one value and only a few observations having a
different value, a subset may have all its observations having the same value
in the variable.  In this case, there would be no estimate for the
corresponding coefficient.  Similarly, a categorical predictor variable
having a rare category may also cause a problem.  Therefore it might be
necessary to pre-screen and remove such predictor variables, or consider the
predictor variables when dividing the data in a manner to ensure feasible
estimation in each subset.  Given these limitations of the divide-and-combine
approach and the simplicity of binning and rounding, one might prefer one of
the latter approaches.

We focused on generic binning and rounding algorithms: equal-quantile
binning, rounding to a decimal place, and rounding to significant digits.
Alternative, and probably more {\it ad hoc}, binning and rounding approaches
might be desirable in certain applications.  For example, one might bin the
outcome with more convenient or interpretable cutoff values, or transform the
outcome to a different scale and then round it, or round the outcome
differently in different regions.

%Left skewedness.

In summary, we have provided three approaches to the problem of fitting CPMs
to big data.  They perform quite well and take a reasonable amount of time
and computer resources to finish.  These approaches have been implemented in
our \verb|cpmBigData| R package. % (part of PResiduals or standalone?)

\section{References}

Zeng D, Lin DY. Maximum likelihood estimation in semiparametric regression
models with censored data. {\it J R Stat Soc Series B (Methodol)}.
2007;69(4):507--564.

Liu Q, Shepherd BE, Li C, Harrell Jr FE. Modeling continuous response
variables using ordinal regression. {\it Statistics in
  Medicine}. 2017;36:4316--4335.

McCullagh P. Regression models for ordinal data. {\it J R Stat Soc Series B
  (Methodol)}. 1980;42(2):109--142.

Agresti A. {\it Analysis of Ordinal Categorical Data}. Second edn. Hoboken,
New Jersey: John Wiley \& Sons; 2010.

Li C, Zeng D, Tian Y, Shepherd BE. A semiparametric transformation model for
continuous outcomes. 2021 (submitted)

Sall J. A monotone regression smoother based on ordinal cumulative logistic
regression. In: ASA Proceedings of Statistical Computing Section, Atlanta,
Georgia; 1991:276--281.

Harrell FE. rms: Regression modeling strategies.
\verb|http://CRAN.R-project.org/package=rms|, R package version 6.0-1; 2020.

Nugteren C, Codreanu V. CLTune: A generic auto-tuner for OpenCL kernels.  In
{\it 2015 IEEE 9th International Symposium on Embedded Multicore/Many-core
  Systems-on-Chip}; 2015.

Ballester-Ripoll R, Paredes EG, Pajarola R. Sobol tensor trains for global
sensitivity analysis.  {\it arXiv Computer Science / Numerical
  Analysis}. 2017;arXiv:1712.00233

\end{document}